\documentclass[conference]{IEEEtran}
\usepackage{amssymb,amsmath}
\usepackage{cite}
\usepackage{graphicx}
\usepackage[caption=false]{subfig}
\usepackage[font=footnotesize]{caption}
\usepackage{psfrag}
\usepackage{url}
\usepackage[latin1]{inputenc}
\usepackage[absolute,overlay]{textpos}
\usepackage{tikz}
\usetikzlibrary{arrows,calc,decorations.markings}
\usetikzlibrary{topaths}
\usetikzlibrary{shapes,trees}
\usetikzlibrary{arrows}
\usetikzlibrary{shadows}
\usetikzlibrary{positioning}
\usetikzlibrary{matrix}
\usetikzlibrary{shapes.geometric}
\usetikzlibrary{decorations.pathmorphing}
\usepgflibrary{patterns}
\usetikzlibrary{calc}
\usetikzlibrary{fit}					

\usepackage{pgf}
\usetikzlibrary{arrows,automata}
\usepackage[latin1]{inputenc}

\usepackage[ruled,linesnumbered]{algorithm2e}

\usepackage{algorithmic}

\usepackage{textcomp}
\usepackage{xcolor}
\def\BibTeX{{\rm B\kern-.05em{\sc i\kern-.025em b}\kern-.08em
    T\kern-.1667em\lower.7ex\hbox{E}\kern-.125emX}}

\usepackage{amsthm}

\usepackage{tabularx}
\usepackage{makecell}
\usepackage{multirow}

\usepackage{amsmath}
\usepackage{amssymb}
\usepackage{bbm}
\usepackage{bm}
\usepackage{comment}
\usepackage[capitalize]{cleveref}

\usepackage{pgfplots}
\usepackage{float}
\usepackage{booktabs}

\pgfplotsset{compat=1.15}

\usepackage[bottom=1 in,top=0.75in,left=0.63in,right=0.63 in]{geometry}

\begin{document}

\title{Meta-Learning Based Few Pilots Demodulation and Interference Cancellation For NOMA Uplink}
\author{
	\IEEEauthorblockN{Hebatalla Issa, Mohammad Shehab and Hirley Alves 
	}
	\IEEEauthorblockA{Centre for Wireless 	Communications (CWC), University of Oulu, Finland \\
	Email: firstname.lastname@oulu.fi}
}
\maketitle

\vspace{0mm}
\begin{abstract}
Non-Orthogonal Multiple Access (NOMA) is at the heart of a paradigm shift towards non-orthogonal communication due to its potential to scale well in massive deployments. Nevertheless, the overhead of channel estimation remains a key challenge in such scenarios. This paper introduces a data-driven, meta-learning-aided NOMA uplink model that minimizes the channel estimation overhead and does not require perfect channel knowledge. Unlike conventional deep learning successive interference cancellation (SICNet), Meta-Learning aided SIC (meta-SICNet) is able to share experience across different devices, facilitating learning for new incoming devices while reducing training overhead. Our results confirm that meta-SICNet outperforms classical SIC and conventional SICNet as it can achieve a lower symbol error rate with fewer pilots. 
\end{abstract}
\section{Introduction}
The path to future wireless networks is encompassed by a massive deployment of devices, enabling smart cities, autonomous vehicles, and many unforeseen scenarios. Machine-type Communication (MTC) services 
have revolutionized the wireless communications industry by shifting the focus toward the Internet of Things (IoT). Besides a large number of devices, these mMTC devices are often battery constraint, with limited computational capability, and have heterogeneous and sporadic traffic patterns. Combined, these characteristics impose many challenges for the design of efficient random-access procedures and radio resource management. 
Therefore, researchers are increasingly studying multiple access techniques that are able to scale well to this massiveness while coping with the existence of interference and scarce spectrum. In this context, Non-Orthogonal Multiple Access (NOMA) with Successive Interference Cancellation (SIC) has shown potential and has been the focus of academy and industry in recent years \cite{NOMA_survey,NOMA_5G, art:yuanCM2020, ShahabCST2020, ElbayoumiCST2020, DOSTI, van2022deep}. This is because NOMA schemes allow for multiple-user transmission with superior performance compared to conventional orthogonal schemes. These recent surveys evince the popularity and potential of NOMA and overview key characteristics, techniques, and applications \cite{NOMA_survey, ShahabCST2020, ElbayoumiCST2020}.

Despite the recent advances, there are many open challenges concerning NOMA in massive IoT deployments such as latency, complexity of receivers, perfect channel knowledge and effective utilization of radio resources. These challenges prompted researchers to look into data-driven approaches to handle multi-user connectivity \cite{van2022deep, EmirWCL2021, SchaufeleEUSIPCO2022, ZhangTWC2022}. We should note that data-driven learning-based communication systems aim to coexist or replace conventional model-based approaches thanks to their low complexity and ability to adapt well to varying channels. Moreover, the learning-based models can be easily customized to any environment, however, at cost of new training instances. For instance, \cite{SchaufeleEUSIPCO2022} argues that although linear processing has been effective in NOMA systems, non-linear processing is sometimes necessary to maintain good performance. The authors propose a neural network architecture that leverages the benefits of both linear and non-linear processing leading to efficient real-time detection performance. The authors in \cite{EmirWCL2021} introduce a deep-learning-based user detection solution (DeepMuD) for the uplink in massive MTC NOMA. The proposed DeepMuD employs an offline-trained Long Short-Term Memory (LSTM)-based network for multi-user detection without the need for perfect channel state information (CSI). Interestingly, authors report that DeepMuD improves error performance compared to conventional detectors and becomes even better as the number of devices increases. 
%
Authors in \cite{van2022deep} introduce a deep learning approach for estimating the symbols named SICNet in the downlink NOMA. Contrary to conventional SIC, SICNet replaces the interference cancellation blocks with deep neural networks (DNNs) to infer the soft information representing the interfering symbols in a data-driven fashion, yielding robustness against changes in the number of users and power allocation. 

More recently, ML-based transceivers have gained lots of interest. For instance, the authors of \cite{ML_MIMO} proposed a meta-SICNet detection scheme for massive MIMO. The results showed that their meta-SICNet outperforms the conventional MMSE detector.  Furthermore, the work in \cite{Simone_trans} investigates learning-based transceivers using joint learning and meta-learning techniques. An important limitation of the autoencoder-based approach to end-to-end training is that training should be generally carried out from scratch for each new channel. Despite the enhanced performance, such models can be trained on an end-to-end basis irrespective of the underlying modulation or multiplexing scheme. However, when relying on conventional training schemes, these models are required to be trained from scratch for each separate transmission condition, which is cumbersome in MTC networks with heterogeneous traffic and radio resources. Therefore, such networks would benefit from a learning scheme that accumulates experience to facilitate learning in new conditions.

In this context, meta-learning was introduced \cite{finn2017model, ChenNOW2023, beckSurveyMetaReinforcementLearning2023}, also known as learning to learn. Often communications networks are expected to operate under a variety of system configurations. Using conventional learning requires the training of a separate model for each system configuration, leading to data and training time and processing costs. Owing to efficiency, we want to train a single model that would perform across all configurations. However, for such joint training, there may not exist a model that is able to perform well. On the other hand, meta-learning uses data from multiple configurations to infer a model class and learning procedure, enabling learning on configurations of interest. For example, few-shot classification aims to infer a learning procedure that trains a classifier under limited training data from each class. To do so, the learning procedure is inferred from meta-training data that quickly train a classifier on meta-training tasks, rather than training a single model to classify across all tasks \cite{finn2017model, ChenNOW2023}. More specifically, in a wireless communication setting, meta-learning outperforms (with respect to symbol error probability) conventional and joint training in IoT scenarios where devices transmit few pilots while adapting to non-linearities and fading. Learning to communicate on noisy or fading channels requires training to be carried out from scratch for every channel, for example, using pilot symbols. In this case, joint training is equivalent to non-coherent transmission. 
In \cite{WuIOTJ2019}, authors address the sensing and fusion problem in massive MTC, which aims to collect and process a large amount of information and extract key features representing the observed process. However, sensing and fusion impose communication overheads and data redundancy to perform a given accuracy. The authors then proposed a meta-learning adaptive sensing and reconstruction framework that leverages prediction error and sensing decisions so to reduce the amount of communication overhead while guaranteeing robustness. Besides, they show that meta-learning-based approaches outperform conventional machine-learning algorithms in terms of convergence rate.

\subsection{Contributions}
In this context, we build upon \cite{EmirWCL2021, van2022deep, park2019learning} and propose a novel few pilot-aided detection mechanism, named meta-SICNet, for NOMA uplink in massive MTC. Herein, as in \cite{van2022deep}, we infer the interfering symbols via deep neural networks. However, we can reduce training over the network by the use of meta-learning, and in addition, as in \cite{park2019learning}, we reduce the number of pilots required for detection. Therefore, our contributions are 
\begin{itemize}
    \item We propose a data-driven approach that applies meta-learning for NOMA uplink transceiver design. 
    
    \item The network model consists of multiple MTC devices transmitting their superimposed signal to the BS. Then, the meta-SICNet framework in the BS learns to recover the transmitted symbols using the meta-training and adaptation approach. 
    \item The meta-SICNet is able to decode symbols with fewer pilots, in the presence of interference, compared to basic SIC and conventional SICNet methods. 
    \item Extensive simulation results elucidate that the proposed meta-SICNet scheme significantly outperforms both classical SIC and conventional SICNet in terms of outage probability and requires a lower number of pilots (down to 2 pilots). 
\end{itemize}

\subsection{Outline} 
The rest of the paper is organized as follows: section~\ref{sec:system} depicts the system model and the problem formulation. Conventional SICNet and the proposed meta-learning-based SICNET solutions are presented in Sections \ref{sec:SIC} and section \ref{sec:results} elucidates the results, and finally, the paper is concluded in Section~\ref{conclusions}.

\section{System Layout and Problem Formulation}\label{sec:system}

\subsection{System Layout}
\begin{figure}[t!]
	\centering
	\includegraphics[width=1\columnwidth]{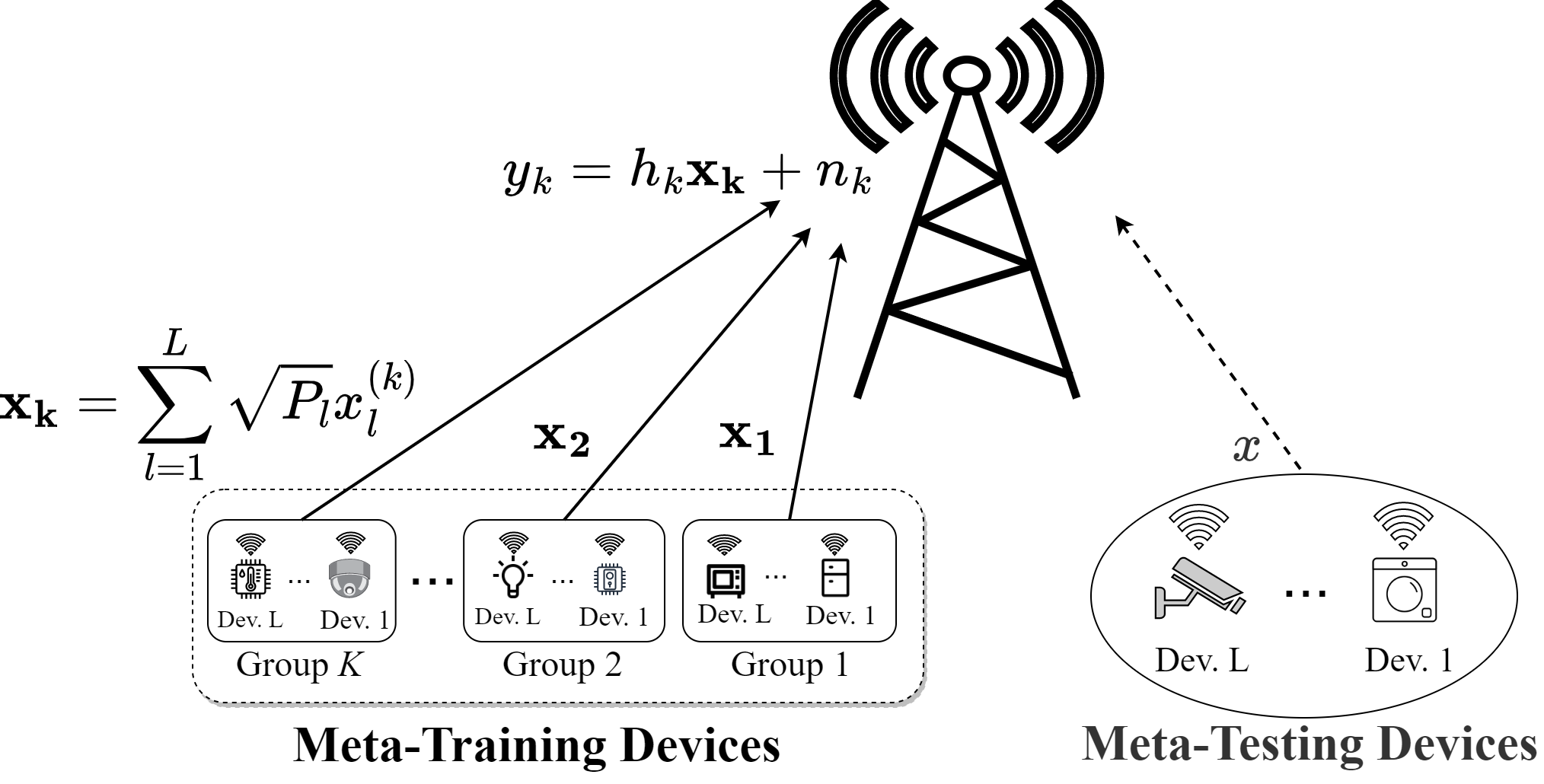}
	\caption{The system model comprises $k$ sets of meta-training devices and one Meta-testing device. Each meta-training set, $k$, transmits a superimposed signal, $\mathbf{x_{k}}$ of $L$-devices to the BS.} \vspace{0mm}
	\label{model}
\end{figure}

We consider a non-orthogonal uplink channel where we have $K$ meta-training device groups and one meta-testing group. For each group $k$, $L$ IoT devices transmit their data to the BS within the same time and frequency resources as illustrated in Fig.\ref{model}. The devices transmit  $\{x_{l=1, \dots, L }^{(k)=1, \dots, K}\}$ symbols to the BS using superposition coding, where the symbol $x_{l}^{(k)}$ is the $l^{th}$ symbol transmitted from the $k^{th}$ device. Specifically, the symbol $x_{l}^{(k)}$ is amplified with the transmitted power $P_l$ for $l=1, \dots, L$. The channel input is the superimposed signal $\mathbf{x_{k}}$ given by
\begin{equation}
   \mathbf{x_{k}} = \sum_{l=1}^{L}  \sqrt{P_{l}}x_{l}^{(k)}
\end{equation}
The symbols are sampled from an $M$-point constellation $\mathcal{S}$, and assumed to be mutually independent with unit mean power, i.e., $\mathop{\mathbb{E}}\left[\left|x^{(k)}_{l}\right|\right]=1$. For simplicity, we assume all devices have the same modulation order $M$.
The channel output at the BS for each $k$ group denoted as $y_{k}$ for $k=1, \dots, K$ is given by 
\begin{equation}
    y_{k} = h_{k} \mathbf{x_{k}} + n_{k},
\end{equation}
where $h_{k} \in \mathop{\mathbb{C}}$ is the channel coefficient between the BS and the device-group $k$, and $n_k  \sim \mathcal{CN}(0, \sigma^2)$ is the complex additive white Gaussian noise (AWGN). 

\subsection{Problem Formulation}
We aim to construct a symbol detection demodulator based on a short packet transmission of a few pilot symbols. For each device $l$ in the device group $k$, the detected symbol, i.e., $\hat{x}_l^{(k)}$, is estimated from the channel output $y_k$. To enable symbol recovery using few-pilot learning, we use a data-driven approach based on meta-learning and DNNs. Assuming no prior information about the channel model at the receiver, the BS can use the signals received from the previous pilot transmissions of $K$ other IoT device groups, which are referred to as meta-training devices and their data as meta-training data. In particular, the BS has available $N$ pairs of pilots $\mathbf{x_{k}}$ and received signal $y_{k}$ for each meta-training group $k = 1,\dots,K$. The meta-training dataset is denoted as $\mathcal{D} = \{D\}_{k = 1,\dots,K}$, where $D_{k} = \{(\mathbf{x_{k}}^{(n)}, y_{k}^{(n)}): n = 1, \dots , N\}$, and $(\mathbf{x_{k}}^{(n)}, y_{k}^{(n)})$ are the pilot-received signal pairs for the $k$th meta-training group of devices.

For the meta-test devices, the BS receives ${P}$ pilot symbols. It collects the ${P}$ pilots received from the
target device in set $\mathcal{D}_{T}\! =\!\! \{(x^{(n)}, y^{(n)})\! :\! n\! =\! 1,\! \dots\!, P\!\}$. The demodulator can be trained using meta-training data $\mathcal{D}$ and the pilot symbols $\mathcal{D}_{T}$ from the meta-test devices.
To recover the symbols, the successive interference cancellation algorithm is utilized, such that the power allocations for the superimposed symbols satisfy $P_1\!\! >\!\! P_2\!\dots\!\! >\! \!P_L$. Hence, the meta-learning-based DNN successively detects each symbol from the channel output signal $y_k$. 
%

\begin{figure}[t!]
	\centering
	\includegraphics[width=1\columnwidth]{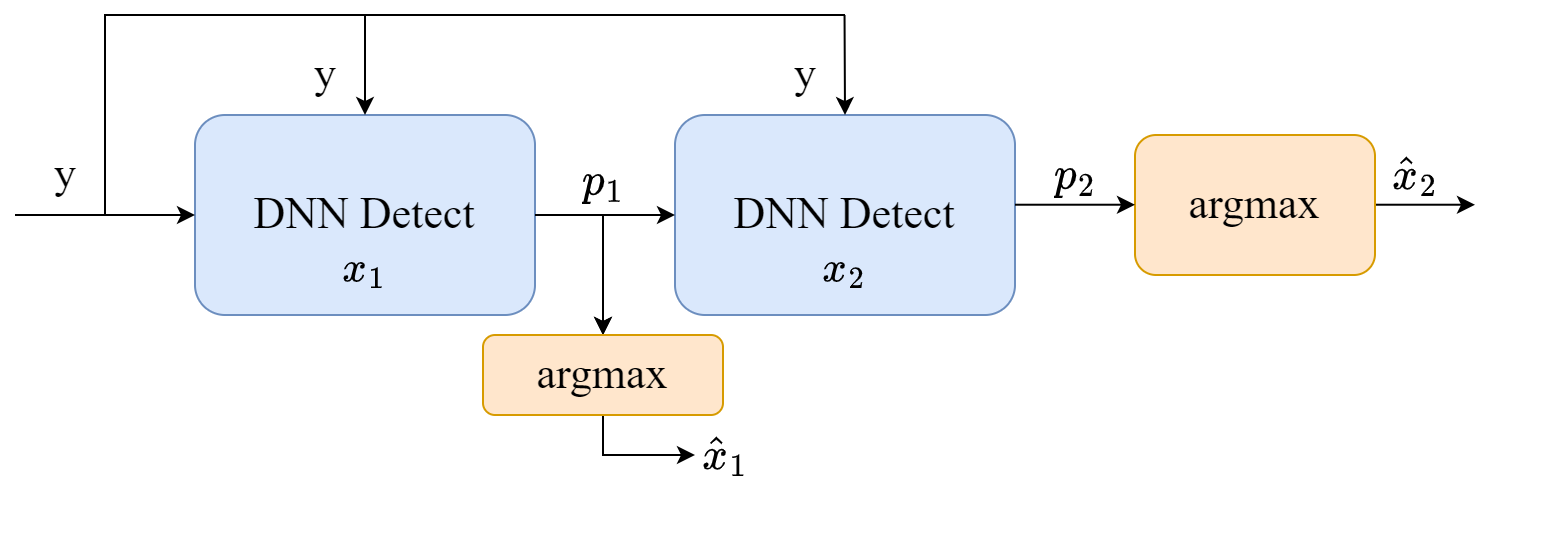}
	\caption{Architecture of SICNet for DNN-based SIC for $L=2$ devices.} 
	\label{sicnet}
\end{figure}

\section{Deep Learning-based SIC} \label{sec:SIC}
In this section, we describe the data-driven solutions based on DNNs to perform SIC and symbol detection. First, we introduce the conventional DNN based SICNet; then we illustrate the proposed meta-SICNet approach. 
\subsection{SICNET}\label{Sicnet}
The DNN-based SIC, called SICNet, was introduced in \cite{van2022deep} to estimate the transmitted symbols in the downlink NOMA users scheme. For the sake of illustration, we consider an uplink scheme with $L=2$ devices in each device group. As depicted in Fig. \ref{sicnet}, the architecture of SICNet is implemented using sequential DNN blocks. Each block performs symbol recovery, which is considered a classification problem. Therefore, the structure of SICNet consists of $L$ DNN stages, where each stage estimates the transmitted symbol of each device. Each stage uses the received signal $y_k$ and the output vector $\mathbf{p}_l$ for $l=1, \dots, L$ that represents the conditional distribution of the corresponding symbol, expressed as 
\begin{align}
    \mathbf{p}_l &= \begin{bmatrix}
    \hat{p}(x_{l}=\gamma_{1}|y_{k}, \mathbf{p}_1, \dots, \mathbf{p}_{l-1}) \\
    \vdots \\
    \hat{p}(x_{l}=\gamma_{M}|y_{k}, \mathbf{p}_1, \dots, \mathbf{p}_{l-1}) \\
    \end{bmatrix},  
\end{align}
where $\gamma_i$ is the $i^{th}$ constellation symbol in the constellation space $\mathcal{S}$ for $i=1,\dots,M$, and $\hat{p}(x_{l}=\gamma_{i}|y_{k}, \mathbf{p}_1, \dots, \mathbf{p}_{l-1})$ is a parametric estimation of the probability of $x_l$ given $y_k$ and the previous estimates $\mathbf{p}_1, \dots, \mathbf{p}_{l-1}$. 



The data-driven feature of the SICNet increases its capability to detect symbols reliably without requiring full knowledge of the channel model. Concretely, it works in a model-agnostic manner contrary to the conventional SIC algorithm \cite{van2022deep}, which requires a restricted channel model \cite{andrews2005interference}. Thus, SICNet operates with arbitrary channel models depending on the classification process of the data-driven DNN. Furthermore, the SICNet architecture only requires knowledge of the modulation order $M$ and the power coefficients of the transmitted signals. 

\subsection{Proposed Meta-Learning Approach}
In this section, we describe the meta-learning approach for demodulation based on \cite{park2019learning}. In our proposed method, the meta-learning-based SICNet allows learning the interference cancellation and signals demodulation using a few-pilot transmission. Moreover, the model can quickly adapt to the change in the channel conditions, in contrast to the classical SICNet. 
\begin{figure}[t!]
	\centering
	\includegraphics[width=0.95\columnwidth]{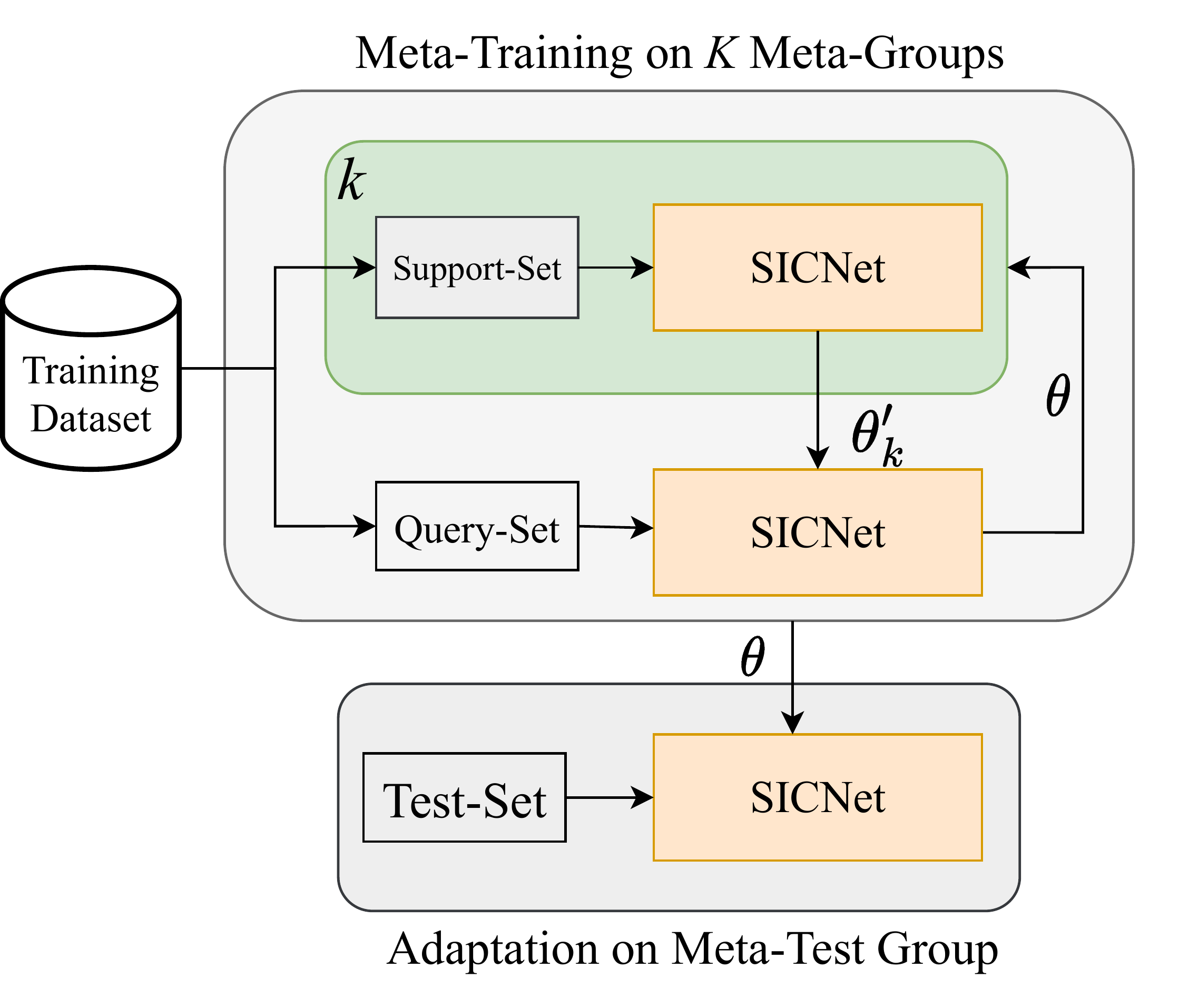}
	\caption{Architecture of meta-SICNet} \vspace{0mm}
	\label{meta-SICNet}
\end{figure}

In our model, the learning is performed in two phases; the meta-learning phase and the test-adaptation phase, as shown in Fig. \ref{meta-SICNet}. In the meta-learning phase, we use the meta-training data $\mathcal{D}$ and iterate over the number of $K$ meta-tasks (meta-devices) to learn a general parameter vector $\theta$ using the inner task parameter $\theta'_{k}$. This can be done offline at the BS while collecting data from training devices. For the meta-test adaptation phase, the learned parameter $\theta$ is adapted to enable fast adaptation based on the few pilots $P$ transmitted by the target (meta-test) devices. Consequently, the test data-set $\mathcal{D}_{T}$ is used to train a demodulator $p_{\theta}(x|y)$ to minimize the cross-entropy loss given by

\begin{equation}
    \centering
    L_{\mathcal{D}_{T}}(\theta) = - \sum_{(x^{(n)}|y^{(n)}) \in \mathcal{D}_{T}} \log p_{\theta}(x^{(n)}|y^{(n)}).
    \label{eq_loss}
\end{equation}
Then, the stochastic gradient descent is used to update the parameter $\theta$ iteratively as 
\begin{equation}
    \centering
    \theta \leftarrow \theta - \eta \nabla_{\theta}\log (x^{(n)}|y^{(n)}),
    \label{eq_update}
\end{equation}
where the pair ${(x^{(n)}|y^{(n)}) \in \mathcal{D}_{T}}$, and $\eta$ is the step size. 
As discussed in \cite{finn2017model}, the purpose of the MAML algorithm is to find the initial parameter $\theta$ such that, for any device, the loss after one iteration of \eqref{eq_update} applied to the received pilots is minimized. The training and testing (adaptation) steps are illustrated in Algorithm \ref{alg1}. 

\RestyleAlgo{ruled}
\begin{algorithm} [!t]
\caption{Meta-Learning based SIC} \label{alg1}
\KwIn{Meta-training data $\mathcal{D} = \{\mathcal{D}_{k=1,\dots,K}\}$, and meta-testing pilot data $\mathcal{D}_{T}$; $N_{tr}^S$ and $N_{tr}^Q$; step size $\alpha$ and $\beta$.}
\KwOut{Learning parameter vector $\mathbf{\theta}$; SER.}  
\vspace{2mm}
 
Randomly initialize the parameter vector $\theta$.\\
\textsc{Start Meta-Training:} \\
\While{not done} {
\For{each meta-training group $k$} {

Randomly split $\mathcal{D}_k$ into two sets; support-set $\mathcal{D}_{k}^{S}$ of size $N_{tr}^S$, and query-set $\mathcal{D}_{k}^{Q}$ of size $N_{tr}^Q$. \\

Calculate the  gradient $\nabla_{\theta}L_{\mathcal{D}_{T}}(\theta)$ from \eqref{eq_loss} with $\mathcal{D}_{T}= \mathcal{D}_{k}^{Q}$, and $\nabla_{\theta}^{2}L_{\mathcal{D}_{T}}(\theta)$. \\

Compute adapted parameters $\theta'$ using  \eqref{eq_update} as
\begin{equation*}
        \centering
    \theta'_k =\theta - \beta \nabla_{\theta}L_{\mathcal{D}_{k}^{Q}}(\theta)
\end{equation*}
}
update the meta-parameter $\theta$:
\begin{equation*}
    \theta \leftarrow \theta - \alpha \nabla_{\theta} \sum_{k=1}^{K} L_{\mathcal{D}_{k}^{S}}(\theta'_k)
\end{equation*}
}
\textsc{Start Meta-Testing Adaptation:} \\
\For{testing epochs}{
Load the learned parameter vector $\theta$.\\
Sample data from $\mathcal{D}_{T}$.\\
Update $\theta$ in the direction of the gradient with step size $\eta$ by 
\begin{equation*}
    \theta \leftarrow \theta - \eta \nabla_{\theta}  L_{\mathcal{D}_{T}}(\theta)
\end{equation*} \\
}
\end{algorithm}

\section{Numerical Analysis} \label{sec:results}
In this section, we evaluate the performance of the proposed meta-SICNet in terms of symbol error rate (SER) and compare it to the performance of classical SIC and conventional SICNet. The simulation parameters for meta-SICNeT are presented in Table \ref{tab:par}. The model considers $K=20$ meta-device groups for training, and each group includes $L=2$ non-orthogonal interference devices for the UL system. The transmitted signal to the BS is modulated using the BPSK modulation scheme of modulation order $2$, where the signal is superimposed by the power coefficients $P_1 = 4\text{ and } P_2=1$. The SICNet architecture comprises 2 DNN blocks, and each DNN consists of 4 layers; the input and output layers and two hidden layers. The number of neurons in each hidden layer is shown in Table \ref{tab:par}. The training loss criteria is based on the \textit{combined loss} from both devices. For the end-to-end training, we used the Adam optimizer with the meta-learning rate $\alpha=0.1$ and inner learning rate $\beta=0.001$. For simplicity, we consider a symmetric channel model of $\pm 1$, where half of the training groups have a channel of $+1$ and the other half $-1$. For adaption on target devices, the channel is chosen randomly as $+1$ or $-1$. The training is performed on an 11-th Gen Intel Core i5 2.40GHz CPU and 16-GB RAM. The training loss converges after 200 epochs.

\begin{table}[t!]
\centering
\caption{meta-SICNet Model Parameters}
\label{tab:par}
\begin{tabular}{|l|c|}
\hline 
\textbf{Parameter}    & \textbf{Value}\\ 
\hline 
Number of device groups for meta-SICNet $K$                   & 20      \\
Number of devices in each group $L$                           & 2       \\
Modulation Number $M$                                         & 2       \\
Power factors of the devices $P_1, P_2$                       & $4, 1$  \\
Number of DNN layers for SICNet                               & 4       \\
Neurons for DNN block 1                                       & 24-12   \\
Neurons for DNN block 2                                       & 32-16   \\
Activation function for hidden layers                         & ReLU    \\
Activation function for output layers                         & Softmax \\
Optimizer                                                     & Adam    \\
Outer learning rate  $\alpha$                                 & 0.1     \\
Inner learning rate  $\beta$                                  & 0.001   \\
Number of training epochs                                     & 300     \\
Number of pilots for training data $N_{tr}^S, N_{tr}^Q$       &  4, 4   \\
Size of testing data for target devices                       & $10^6$  \\
Training SNR                                                  & 6 dB    \\
Learning-rate for adaptation on test devices $\eta$           & 0.001   \\
Number of adaptation epochs                                   &1000     \\
\hline
\end{tabular}
\end{table}

\begin{figure}[t!]
    \centering
    \includegraphics[width=0.99\columnwidth]{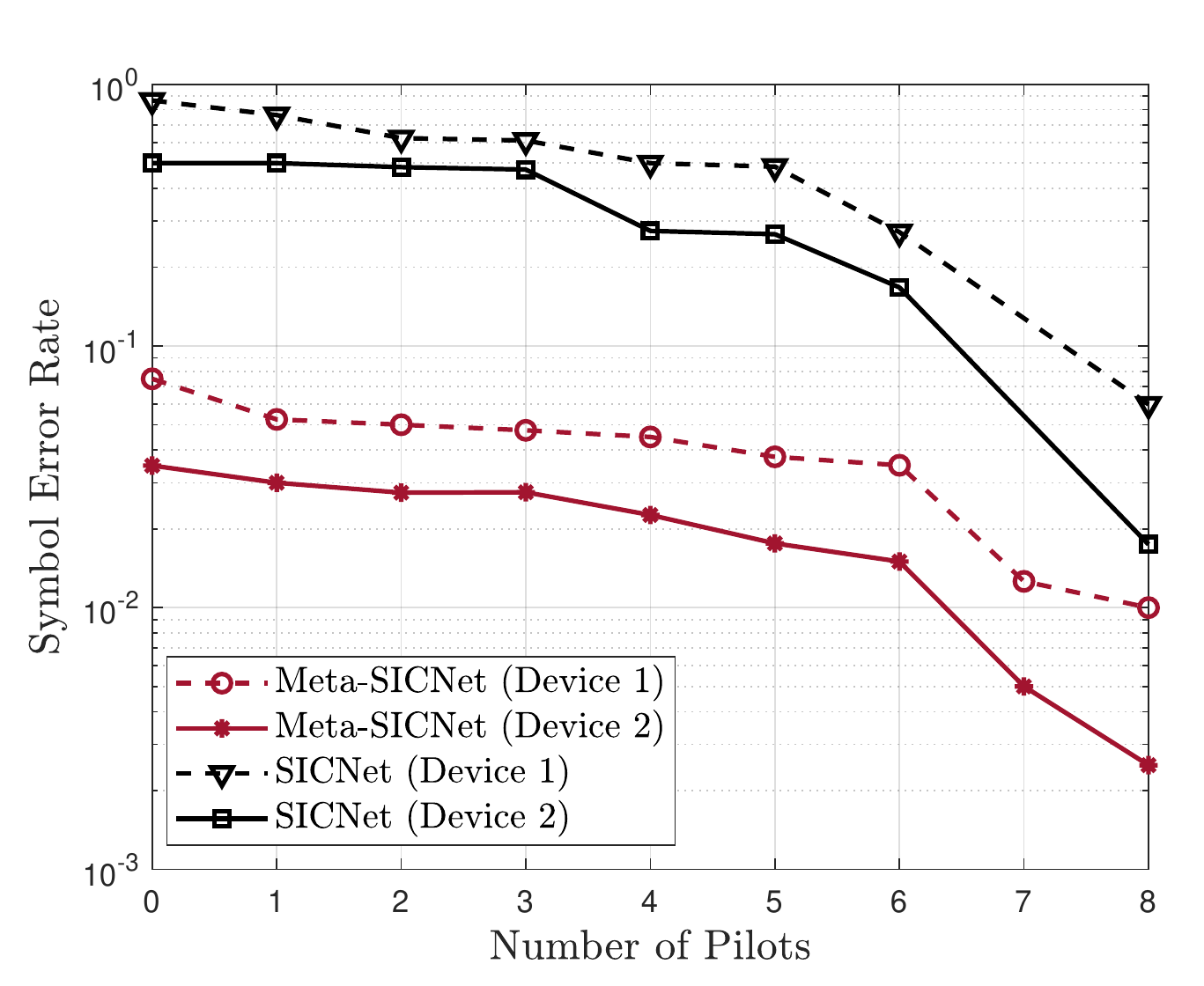}
    \caption{Symbol error rate vs. number of pilots for Meta-testing devices and SICNet (SNR = 15 dB).}
    \label{fig:error_pilots}
\end{figure}

Fig. \ref{fig:error_pilots} depicts the performance of SER for both meta-SICNet and SICNet for a different number of pilots. The simulation was performed for $N_{te} = 1, 2, \dots, 8$ with an SNR value of 15 dB. The results show that the meta-learning approach outperforms the SICNet in terms of SER. Moreover, it is clear that increasing the number of pilots improves learning performance. However, for meta-SICNet, the SER relatively saturates faster at 2 or 3 pilots. Note that device 2 always performs better since interference is already removed after decoding the signal from device 1, which captures the SIC effect.

\begin{figure}[t!]
    \centering
    \includegraphics[width=1\columnwidth]{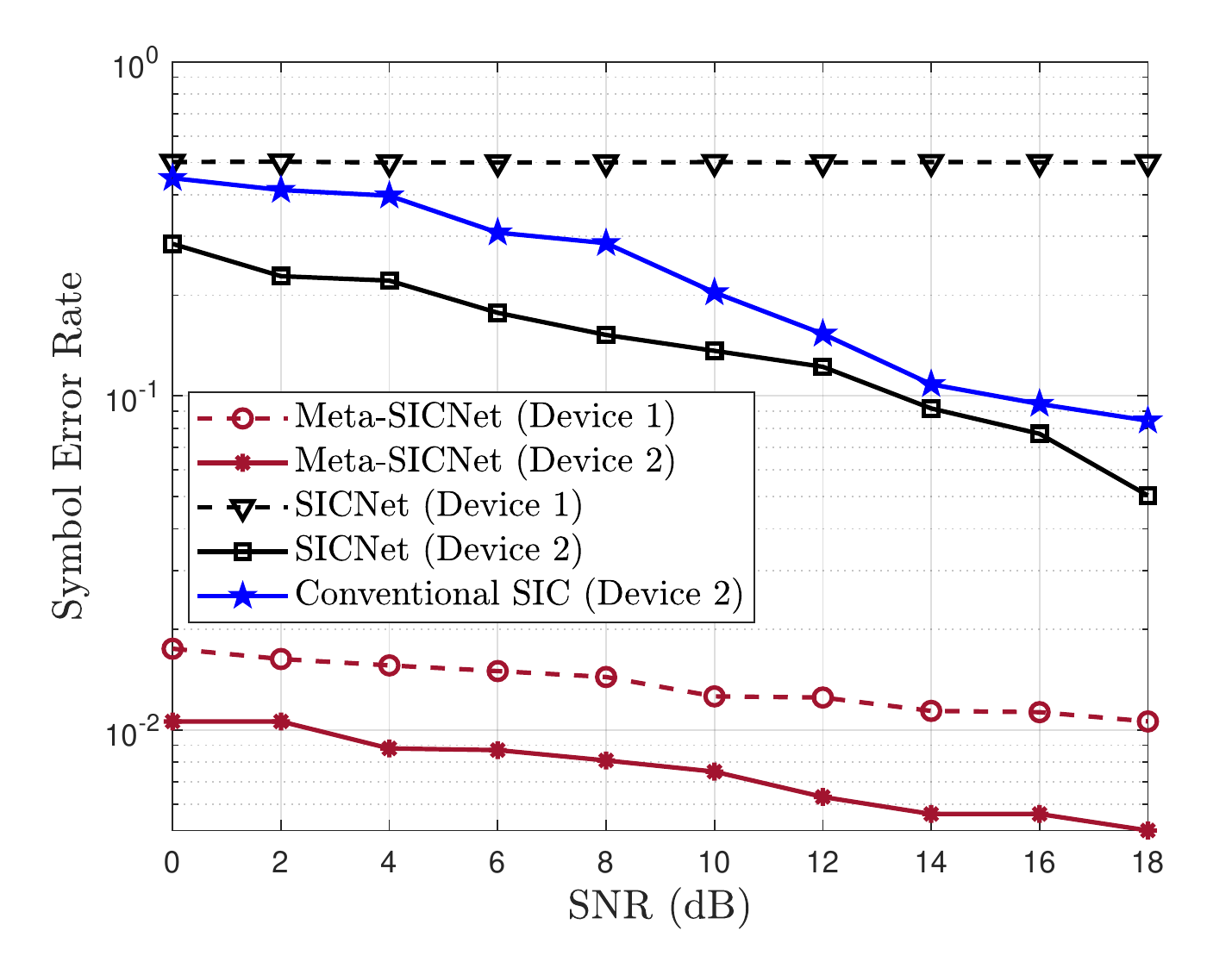}
    \caption{Symbol error rate vs. SNR for Meta-testing devices, SICNet, and conventional SIC (number of pilots = 4).}
    \label{fig:error_snr}
\end{figure}

In Fig. \ref{fig:error_snr}, we plot the symbol outage probability as a function of the SNR for SNR values $\{0, 2, \dots, 18\}$ using 4 pilot symbols. It appears that the meta-SICNet can capture the effect of SNR as the symbol error rate significantly improves when increasing the SNR. Again, it is obvious that the meta-SICNet outperforms conventional SICNet and the classical SIC for all SNR values, which proves the superiority of the proposed meta-learning solution for in NOMA uplink. Note that the classical SIC is the worst performer here, so we just added one curve for illustration.

\begin{figure}[t!]
    \centering
    \includegraphics[width=0.99\columnwidth]{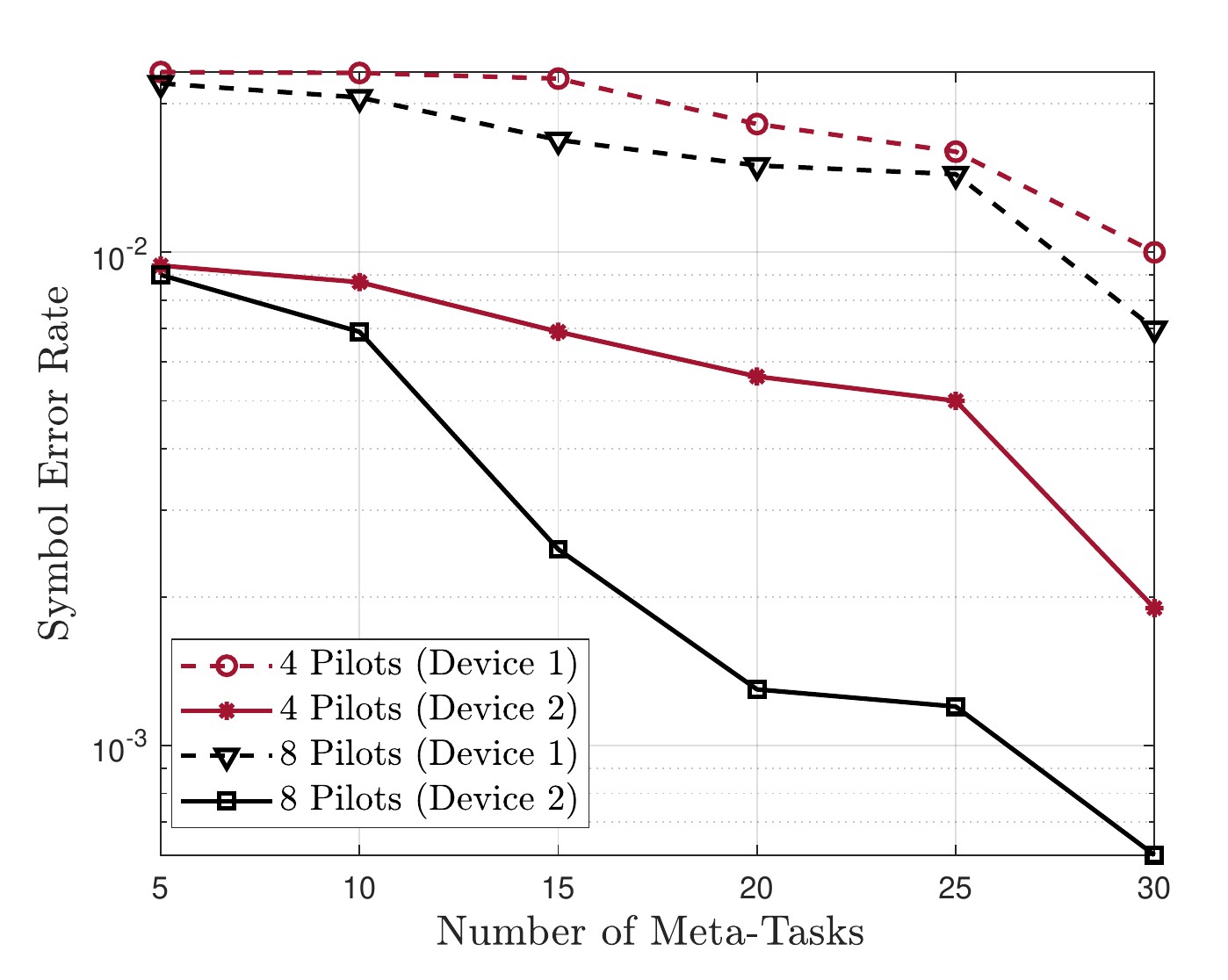}
    \caption{Symbol error rate vs. number of meta-training tasks (number of device groups) for SNR = 15 dB, and number of pilots {4, 8}.}
    \label{fig:error_ntasks}
\end{figure}

In Fig. \ref{fig:error_ntasks}, we address the performance of meta-SICNet for a different number of meta-training tasks (device groups) with 15dB SNR and $N_{te}=\{4,8\}$ pilots. The simulation shows that the capacity of learning for the meta-SICNet model improves when increasing the meta-training device groups. Therefore, the model can achieve lower SER if the training experience is shared between more device groups.

Table \ref{tab:comp} illustrates the training complexity for both meta-SICNet and SICNet in terms of the number of parameters, training time per epoch, and adaptation (testing) time. Although the meta-SICNet approach achieves higher computational complexity, the training can be done offline among the meta-training device groups. Then, the online adaptation process is performed to continuously update the model parameters without having to train the model again. On contrary, the SICNet model requires performing the whole training process online whenever the channel condition alters. Therefore, the testing (adaptation) time for the meta-based approach is much less than SICNet for the same number of pilots. Hence, meta-SICNet can achieve lower SER with relatively low online complexity (i.e, only 4.01 ms of adaptation time).

\begin{table}[t!]
\centering
\caption{Complexity analysis of meta-SICNet vs. SICNet, 8 pilots}
\label{tab:comp}
\begin{tabular}{c|c|c|c}
\toprule 
\textbf{Model} &\textbf{\# Parameters} & \textbf{Training time} & \textbf{Test-time}\\ 
\midrule
\textbf{Meta-SICNet}             & 2240 & 50.3 ms & 4.01 ms  \\

\textbf{SICNet}                  & 1120 & 10.9 ms & 41.01 ms  \\
\bottomrule
\end{tabular}
\end{table}

\section{Conclusions}\label{conclusions}
In this work, we introduced a data-driven, meta-learning-aided NOMA uplink model. Unlike classical SIC and conventional deep learning SIC, the proposed 
meta-SICNet can accumulate experience across different devices, facilitating the learning process for newly introduced devices and reducing the training overhead. Our results confirm that meta-SICNet outperforms conventional SICNet as it can achieve a lower symbol outage probability. Moreover, meta-SICNet can converge faster and renders significantly good performance for a few pilots (only 2 or 3 pilots). Although meta-SICNet consumes a longer time for training, the training phase is performed offline, and the online adaptation phase consumes a very short time ($\approx$ 4 ms) compared to conventional SIC, which reduces the online complexity. There are plenty of possible extensions to the meta-SICNet approach proposed here, among which full-duplex NOMA with self-interference cancellation, adaptation to a higher number of devices, and meta-learning aided massive MIMO.

\section*{Acknowledgments} \vspace{1mm}
This work is partially supported by the Academy of Finland, 6G Flagship program (Grant no. 346208) and FIREMAN (Grant no. 326301), and the European Commission through the Horizon Europe project Hexa-X (Grant Agreement no. 101015956).

\bibliographystyle{IEEEtran}
\bibliography{references}
\end{document}